\documentclass[%
reprint,
superscriptaddress,
amsmath,amssymb,
aps,
prb,
longbibliography,
]{revtex4-1}

\usepackage{pdfpages}
\usepackage{mhchem}
\usepackage{graphicx}
\usepackage{dcolumn}
\usepackage{bm}
\usepackage{hyperref}
\hypersetup{
colorlinks = true,   
urlcolor = blue,    
linkcolor = magenta,   
citecolor = blue  
}

\usepackage{float}
\usepackage{color}

\usepackage{soul}
\setstcolor{red}

\makeatletter
\AtBeginDocument{\let\LS@rot\@undefined}
\makeatother

\begin{document}
\preprint{APS/123-QED}
\title{Majorana corner flat bands in two-dimensional second-order topological superconductors}
\author{Majid Kheirkhah}
\email{kheirkhah@ualberta.ca}
\affiliation{Department of Physics, University of Alberta, Edmonton, Alberta, Canada T6G 2E1}

\author{Yuki Nagai}
\affiliation{CCSE, Japan Atomic Energy Agency, 178-4-4, Wakashiba, Kashiwa, Chiba, 277-0871, Japan} 
\affiliation{Mathematical Science Team, RIKEN Center for Advanced Intelligence Project (AIP), 1-4-1 Nihonbashi, Chuo-ku, Tokyo 103-0027, Japan}

\author{Chun Chen}
\affiliation{Department of Physics, University of Alberta, Edmonton, Alberta, Canada T6G 2E1}

\author{Frank Marsiglio}
\affiliation{Department of Physics, University of Alberta, Edmonton, Alberta, Canada T6G 2E1}
\date{\today}
\begin{abstract}  
In this paper we find that confining a second-order topological superconductor with a harmonic potential leads to a proliferation of Majorana corner modes. As a consequence, this results in the formation of Majorana corner flat bands which have a fundamentally different origin from that of the conventional mechanism. This is due to the fact that they arise solely from the one-dimensional gapped boundary states of the hybrid system that become gapless without the bulk gap closing under the increase of the trapping potential magnitude. The Majorana corner states are found to be robust against the strength of the harmonic trap and the transition from Majorana corner states to Majorana flat bands is merely a smooth crossover. As a harmonic trap can potentially be realized in heterostructures, this proposal paves a way to observe these Majorana corner flat bands in an experimental context.
\end{abstract}

\pacs{Valid PACS appear here}
\maketitle
\section{Introduction}
Topological insulators (TIs) and superconductors (TSCs) have opened a new avenue for topology in both condensed matter and high-energy physics \cite{qi2011topological,hasan2010colloquium,moore2010birth, sato2017topological,qi2009time, qi2010topological,zhang2013time,wong2012majorana,sato2010topological,zhang2013topological, haim2016no}. These topological materials have a bulk-boundary correspondence: there are gapped bulk states characterized by a topological invariant and topologically protected gapless states localized on boundaries. Recently, a new concept of topological materials, so-called higher-order TSCs and TIs, have attracted much attention \cite{benalcazar2017quantized,benalcazar2017electric, schindler2018higher,langbehn2017reflection,ezawa2018higher, khalaf2018higher,geier2018second,imhof2018topolectrical, franca2018anomalous, cualuguaru2019higher}. 
In terms of the conventional understanding of the bulk-boundary correspondence, these materials are topologically trivial, since both bulk and boundary states are gapped. 
However, there are robust gapless states on the ``edge" of the boundary of these materials. In $d$-spatial dimensions, $n$-th order topological insulators and superconductors have $(d-n)$-dimensional gapless localized modes. Therefore, higher-order TSCs are good candidate systems for finding Majorana zero modes (MZMs) because they are localized as ($d-n$)-dimensional bound states. The MZMs are their own antiparticles and obey non-Abelian statistics \cite{read2000paired,volovik1999fermion, kitaev2001unpaired, alicea2012new,lian2018topological}. Conventional TIs and TSCs are understood as examples of first order topological materials.

In second-order TSCs these MZMs have been studied at the corners of a two-dimensional (2D) system and hinges of a three-dimensional (3D) system where neighboring hinges have different chiralities \cite{schindler2018higher_bis, PhysRevB.100.020509}. These zero-energy corner modes are known as Majorana corner states (MCSs) which have been studied in various kinds of system such as high-temperature superconductors (SCs) \cite{lee2006doping, sigrist1991phenomenological, wu2019higher, yan2018majorana, wang2018high, hsu2018majorana}, $s$-wave superfluid \cite{PhysRevLett.123.060402, PhysRevB.100.020509}, systems with an external magnetic field \cite{liu2018majorana, volpez2019second, zhu2018tunable}, and 2D and 3D second-order TSCs \cite{wang2018weak}.

\begin{figure}[t]
\centering
\includegraphics[scale =0.6]{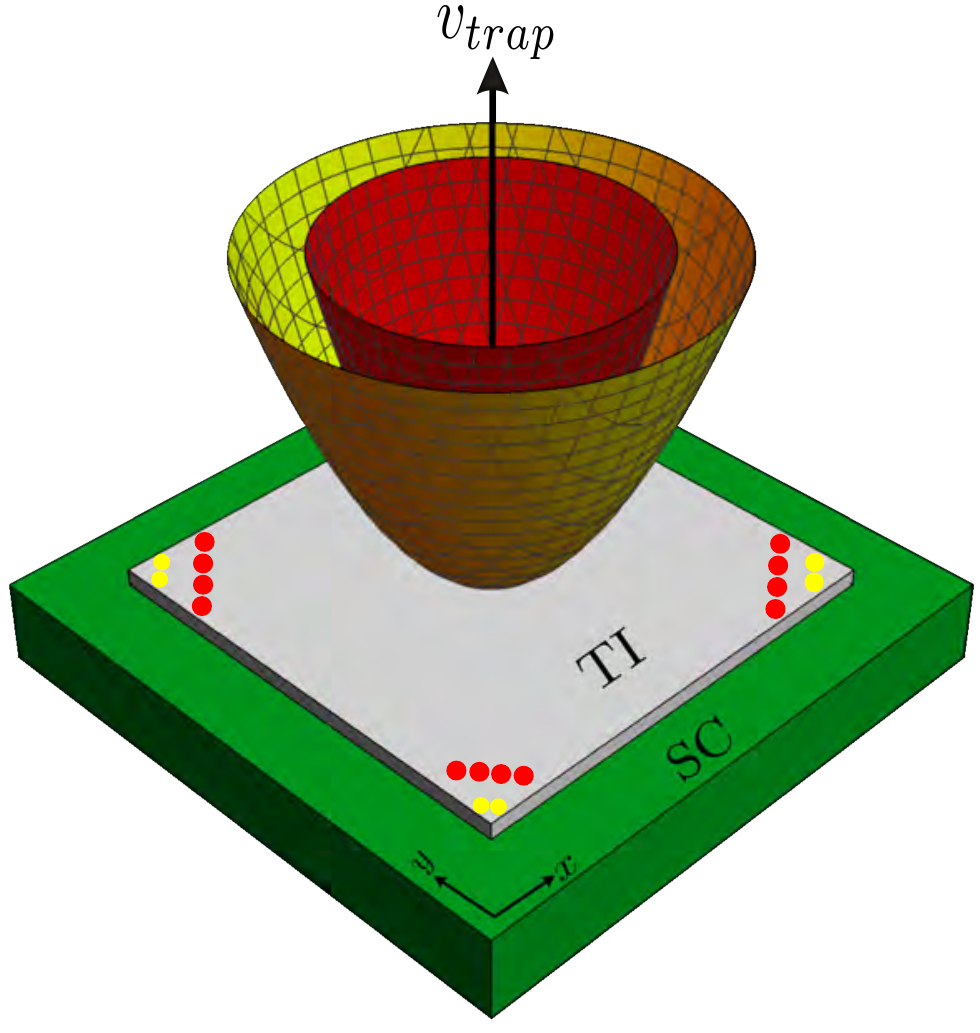}
\caption{(Color online) Schematic picture of a 2D TI approximated by a high-temperature $d$-wave SC in the presence of two different 2D harmonic potentials, where the red paraboloid has larger potential magnitude than the yellow one. The Majorana zero modes are shown by red and yellow points. The number of MCSs (yellow points) are gradually increases as the HP magnitude increases which results in the formation of Majorana corner flat bands (red points).}
\label{schematic}
\end{figure}

An important issue is to both find and determine the robustness of the MCSs. As shown in Ref.~\cite{yan2018majorana}, the MCSs are robust in a 2D TI, also know as the quantum spin Hall insulator, in proximity with cuprate-based \cite{zareapour2012proximity,li2015realizing,wang2013fully}, or iron-based \cite{stewart2011superconductivity,xu2016topological,wang2018evidence,zhang2018observation} high-temperature SCs, as shown schematically in Fig.~\ref{schematic}.
In this system, modest edge imperfections do not affect the MCSs while big edge imperfections just create new corners that host their own MCSs.
This study suggests that the shape of the system is important and it has some effects on the MCSs.

Here, we consider the harmonic potential (HP) as an example of a gradual confining potential. The HP is defined as $V_i = v_{\rm{trap}} \big\{ (i_x-i^c_x)^2 + (i_y-i^c_y)^2 \big\}$ where $v_{\rm{trap}}$ is the potential magnitude, and $i^c = (i^c_x,i^c_y)$ is the coordinate of the central site of the square lattice which we take as the origin. There is no well-defined edge associated with a gradual confining potential and near a MCS, there is no additional corner where additional MCSs can appear. In the limit of a HP whose magnitude is much larger than the insulating gap, the region near the four corners of 2D TI in proximity with a $d$-wave SC becomes the location for unconventional superconductivity. It should be pointed out that in the homogeneous system by increasing the chemical potential there is a topological phase transition with bulk gap closing. Moreover, a time-reversal invariant (TRI) $d$-wave SC with [110] surface, has zero-energy flat dispersion due to a nontrivial one-dimensional (1D) winding number at a large enough chemical potential \cite{sato2011topology, wong2012majorana, wong2013majorana, nagai2017time}. Therefore, there might be a transition from MCSs to Majorana flat bands (MFBs) in a nonhomogeneous system with increasing the HP magnitude. The question arises: to what degree can the number of MCSs be varied as we change the magnitude of a confining potential? Will a topological phase transition occur as the bulk gap closes, or will there be a crossover as the bulk gap continues to decrease without closing?

In this paper, we show that in a second-order TSCs with a confining HP the MCSs-MFBs transition is merely a crossover: the number of the MZMs gradually increases while there is no 2D bulk gap closing as the HP magnitude increases. We find that the increase of the number of MZMs indicates the appearance of new Majorana states originating from the fact that {\it only} 1D gapped boundary modes become gapless {\it without} closing the bulk gap. This eventually leads to a new kind of MFBs which we call {\it Majorana corner flat bands (MCFBs)}. It is found that 2D bulk states do not become gapless with increasing the HP magnitude in a system with open-boundary conditions (OBCs), {\it i.e.}, the host system for the MCSs. We show that the crossover behavior is important for explaining the increase in the number of MZMs. In contrast, in a system with periodic boundary conditions (PBCs), the 2D bulk gap is closed if the potential magnitude at the corners, $v_{\rm{trap}}(r_c)$, becomes larger than the insulating gap. We confirm that, under sharp potentials such as circular potential, the 2D bulk gap is closed in both OBCs and PBCs systems.

\begin{figure}[t]
\centering
\includegraphics[scale = 0.33]{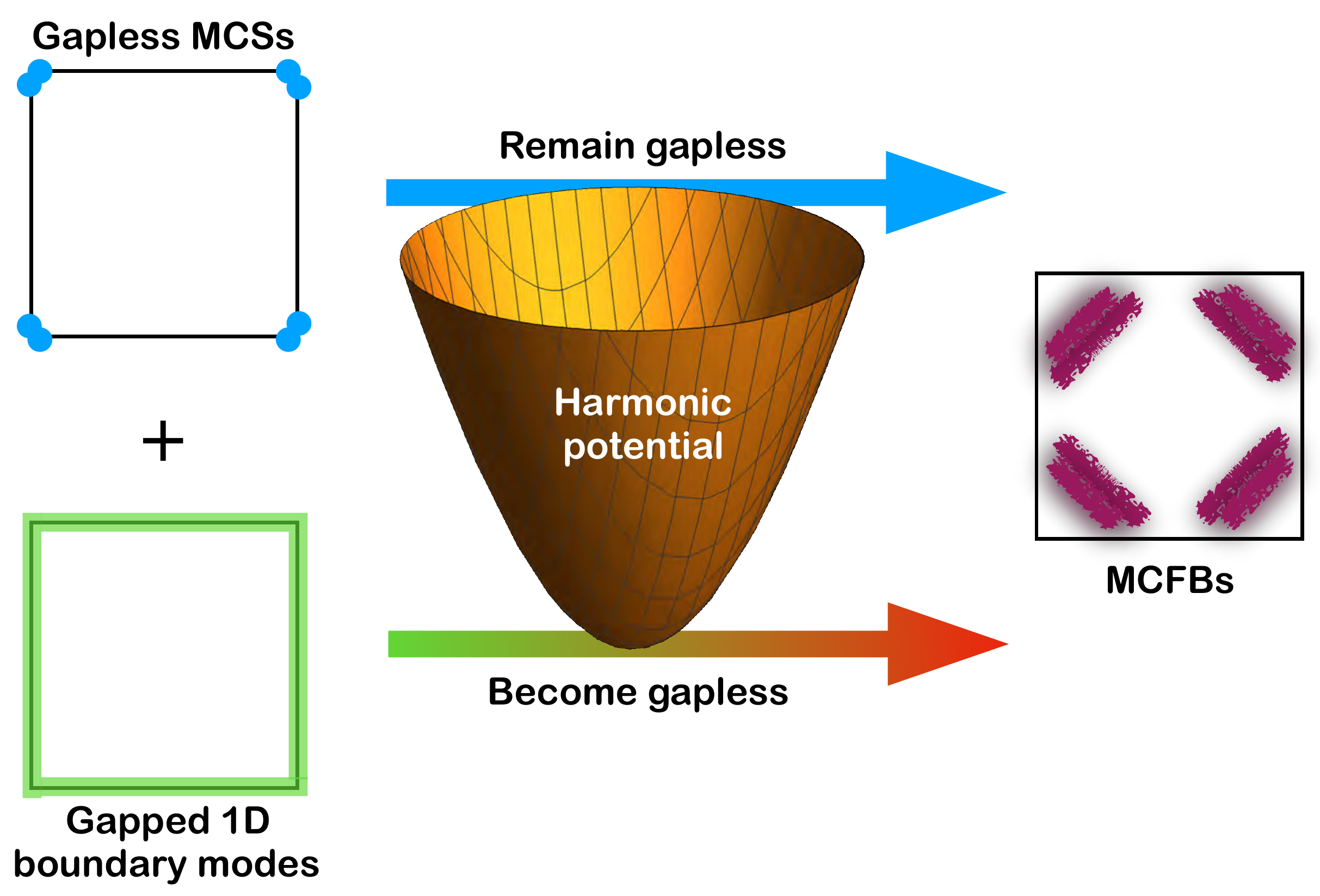}
\caption{(Color online) A schematic picture of the procedure by which MCSs become MCFBs. (Left panel) There is no HP. The system has topologically distinct 1D gapped modes on the adjacent boundaries which leads to gapless MCSs at the intersection of two different boundaries, {\it i.e.}, corners. (Right panel) After including the HP, the gapless MCSs remain gapless. However, the 1D gapped boundary modes become gapless modes.}
\label{sche_2}
\end{figure}

This paper is organized as follows. We introduce the model of a two-dimensional TI with a proximity-induced $d$-wave superconducting gap, as shown schematically in Fig.~\ref{schematic}. This system hosts the MCSs when the chemical potential is smaller than the insulating gap \cite{yan2018majorana}. Then, it will be shown that the MCSs-MCFBs transition in OBCs system is a crossover and there is no bulk gap closing as the HP magnitude increases.
It will be proposed that this kind of MFB is new and different from the conventional one, which has already been reported along the [110] surface of a TRI 2D nodal $d$-wave SC \cite{sato2011topology}, in the sense that they only originate from MCSs or 1D gapped boundary states as illustrated in Fig.~\ref{sche_2}. Finally, we propose an experimental realization for observing this kind of MFB.
\section{Formalism}
The Bogoliubov-de Gennes (BdG) Hamiltonian of a 2D TI in proximity with a SC is given by,
\begin{equation}
\mathcal{H} = \frac{1}{2} \sum_{ij}
\bm{\psi}^{\dagger}_i
\begin{pmatrix}
 \hat{H}^{\rm N}_{ij} && \hat{\Delta}_{ij} \\ \\
 \hat{\Delta}_{ij}^{\dagger} &&  -\hat{H}^{\rm N \ast}_{ij}
\end{pmatrix}
\bm{\psi}_j,
\label{hami_1}
\end{equation}
with $\bm{\psi}^{\dagger}_i=(
c^{\dagger}_{i,a,\uparrow},
c^{\dagger}_{i,b,\uparrow},
c^{\dagger}_{i,a,\downarrow},
c^{\dagger}_{i,b,\downarrow},
c_{i,a,\uparrow},
c_{i,b,\uparrow},
c_{i,a,\downarrow},
c_{i,b,\downarrow})$.
Here, $c^{\dagger}_{i,\alpha,\beta} (c_{i,\alpha,\beta})$
denotes the creation (annihilation) operator of an electron at site $i = (i_x,i_y)$, in orbital $\alpha$ = $a$ or $b$, with spin $\beta$ = $\uparrow$ or $\downarrow$. The normal state and superconducting parts of the Hamiltonian are given by $\hat{H}^{\rm N}_{ij} = m_{ij} \sigma_z s_0 + A^x_{ij} \sigma_x s_z + A^y_{ij} \sigma_y s_0 - \delta_{ij} (\mu - V_i ) \sigma_0 s_0$, and $\hat{\Delta}_{ij}= -\text{i}\Delta_{ij} \sigma_0 s_y$, respectively. The mass term is $m_{ij} = m_0 \delta_{i j} - (t/2) (\delta_{i, i+\hat{x}} + \delta_{i, i - \hat{x}}+ \delta_{i, i+\hat{y}} + \delta_{i, i-\hat{y}})$ where $m_0$ is on-site orbital-dependent energy, and $t$ is the intra-orbital nearest neighbor hopping magnitude along the $x$ and $y$ directions with unit vectors $\hat{x}$ and $\hat{y}$. The $\sigma_{x,y,z}$ and $s_{x,y,z}$ are Pauli matrices acting on orbital and spin degree of freedoms, respectively, and $\sigma_0$ and $s_0$ are $2 \times 2$ unit matrices. Moreover, $A^x_{ij} = (-\text{i} \lambda /2) (\delta_{i, i+\hat{x}} - \delta_{i,i-\hat{x}})$,  $A^y_{ij} = (-\text{i} \lambda /2) (\delta_{i, i+\hat{y}} - \delta_{i,i-\hat{y}})$ are the spin-orbit coupling terms with magnitude $\lambda$, where the symbol $\text{i}$ denoting $\sqrt{-1}$ should not be confused with the site-index that generally occurs subscripted. The chemical potential is given by $\mu$ and $V_i$ is the 2D single particle potential magnitude at site $i$. The $d$-wave superconductivity order parameter with $d_{x^2 - y^2}$ symmetry is expressed as $\Delta_{ij} = (\Delta_d/2)(\delta_{i, i+\hat{x}} + \delta_{i, i - \hat{x}} -\delta_{i, i+\hat{y}} -\delta_{i, i - \hat{y}})$.

In the absence of any single particle potential ($V_i = 0$), the corresponding BdG Hamiltonian in momentum space\cite{remark1} can be written as $\mathcal{H} = \sum_{\bm{k}} \Psi_{\bm{k}}^{\dagger} \mathcal{H}_{\text{BdG}} \Psi_{\bm{k}}/2$ where $\Psi_{\bm{k}}^{\dagger} =
(c^{\dagger}_{\bm{k},a,\uparrow},
c^{\dagger}_{\bm{k},b,\uparrow},
c^{\dagger}_{\bm{k},a,\downarrow},
c^{\dagger}_{\bm{k},b,\downarrow},
c_{-\bm{k},a,\uparrow},
c_{-\bm{k},b,\uparrow},
c_{-\bm{k},a,\downarrow},
c_{-\bm{k},b,\downarrow})$ 
and,
\begin{align}
\mathcal{H}_{\text{BdG}} = &~ m(\bm{k}) \sigma_z s_{0} \tau_z 
+ \lambda \sin k_x \sigma_x s_z \tau_{0} 
+ \lambda \sin k_y \sigma_y s_{0} \tau_z
\nonumber \\ &
+ \Delta(\bm{k}) \sigma_{0} s_y \tau_y
-\mu \sigma_{0} s_{0} \tau_z,
\end{align} 
where $\tau_{x,y,z}$ are Pauli matrices in particle-hole space, and $\tau_0$  is the $2 \times 2$ unit matrix, and
\begin{align}
&\Delta(\bm{k}) = \Delta_d( \cos k_x - \cos k_y),
\\&
m(\bm{k}) = m_{0} -t(\cos k_x + \cos k_y).
\end{align}
The BdG Hamiltonian possesses both particle-hole symmetry (PHS), {\it i.e.}, $\mathcal{P} \mathcal{H}^{\ast}_{\text{BdG}}(\bm{k}) \mathcal{P}^{-1} = -\mathcal{H}_{\text{BdG}}(-\bm{k})$ where $\mathcal{P} = \sigma_{0} s_0 \tau_x$, and time-reversal symmetry (TRS), {\it i.e.}, $\mathcal{T} \mathcal{H}^{\ast}_{\text{BdG}}(\bm{k}) \mathcal{T}^{-1} = \mathcal{H}_{\text{BdG}}(-\bm{k})$ where $\mathcal{T} = i \sigma_{0} s_y \tau_0$. The combination of these two anti-unitary symmetries gives rise to a unitary chiral symmetry, {\it i.e.}, $\mathcal{S} \mathcal{H}_{\text{BdG}}(\bm{k}) \mathcal{S}^{-1} = -\mathcal{H}_{\text{BdG}}(\bm{k})$
where $\mathcal{S} = -i \mathcal{P} \mathcal{T}$. The intrinsic particle-hole symmetry of the BdG Hamiltonian guarantees that the eigenstates whose eigenvalues are zero satisfy the Majorana conditions as discussed in the Appendix. It should be noted that the above Hamiltonian without $d$-wave pairing becomes the paradigmatic Bernevig-Hughes-Zhang (BHZ) model of 2D TIs \cite{bernevig2006quantum}. In addition, the normal state dispersion is given as
\begin{equation}
E(\bm{k}) = \pm \sqrt{m^2(\bm{k}) + \lambda^2( \sin^2 k_x + \sin^2 k_y)} - \mu,
\end{equation}
so, the insulating gap as an important quantity in our analysis, is given by $|m(\bm{k}=0)| = |m_0 - 2t|$. Through this work, we set $m_0 = 1.5$, $\mu =0$, $t= \lambda = 1$, and $\Delta_d =0.5$. 
\section{Results}
\subsection{Robustness of MCSs and crossover with MCFBs}
In a TRI system, MCSs manifest themselves as Majorana Kramers pairs which are protected zero-energy modes localized at the four corners of the 2D TI in proximity with a $d$-wave. By adding proximity-induced $d$-wave superconductivity, the helical gapless boundary modes of the 2D TI become gapped, so in our system as a 2D square sample, there are 1D gapped localized modes along the four boundaries when the chemical potential is smaller than the insulating gap. By constructing the low-energy Hamiltonian and through the use of localized modes, we can define the topological invariant, which depends on the direction of a boundary \cite{yan2018majorana}. There should be gapless zero modes at four intersections or corners of the 2D TI known as MCSs Since the 1D effective systems along the vertical and horizontal boundaries have different topological invariants or Dirac masses, 
\begin{figure}[!th]
\centering
\includegraphics[scale =0.38]{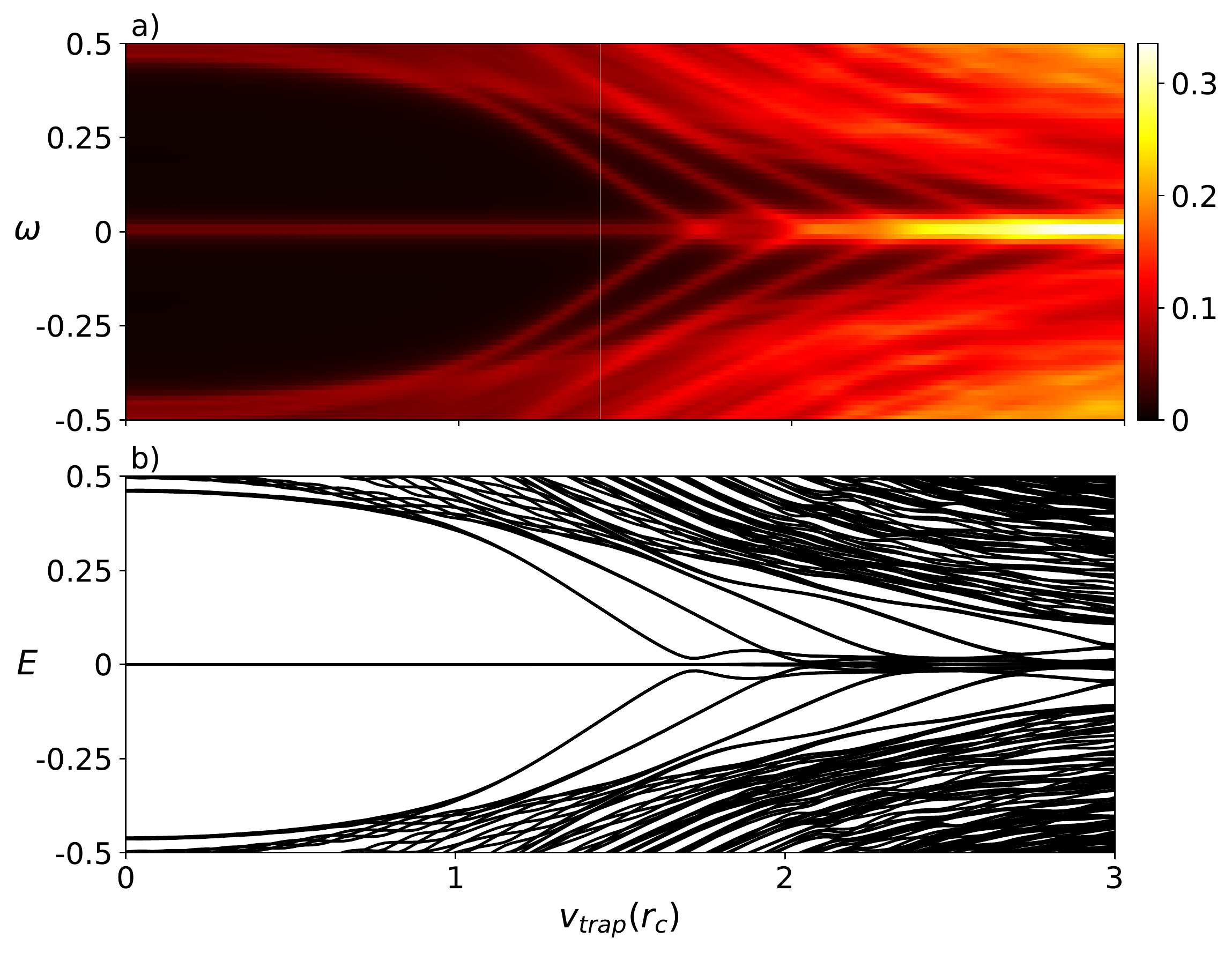}
\caption{(Color online) (a) Density of states, and (b) eigenvalues of the BdG Hamiltonian as a function of $v_{\rm{trap}}(r_c)$ for a $51 \times 51$ square lattice with OBCs. The unit of all values is $\vert m(\bm{k}=0) \vert$.}
\label{fig:obc}
\end{figure}
\begin{figure}[!th]
\centering
\includegraphics[scale =0.38]{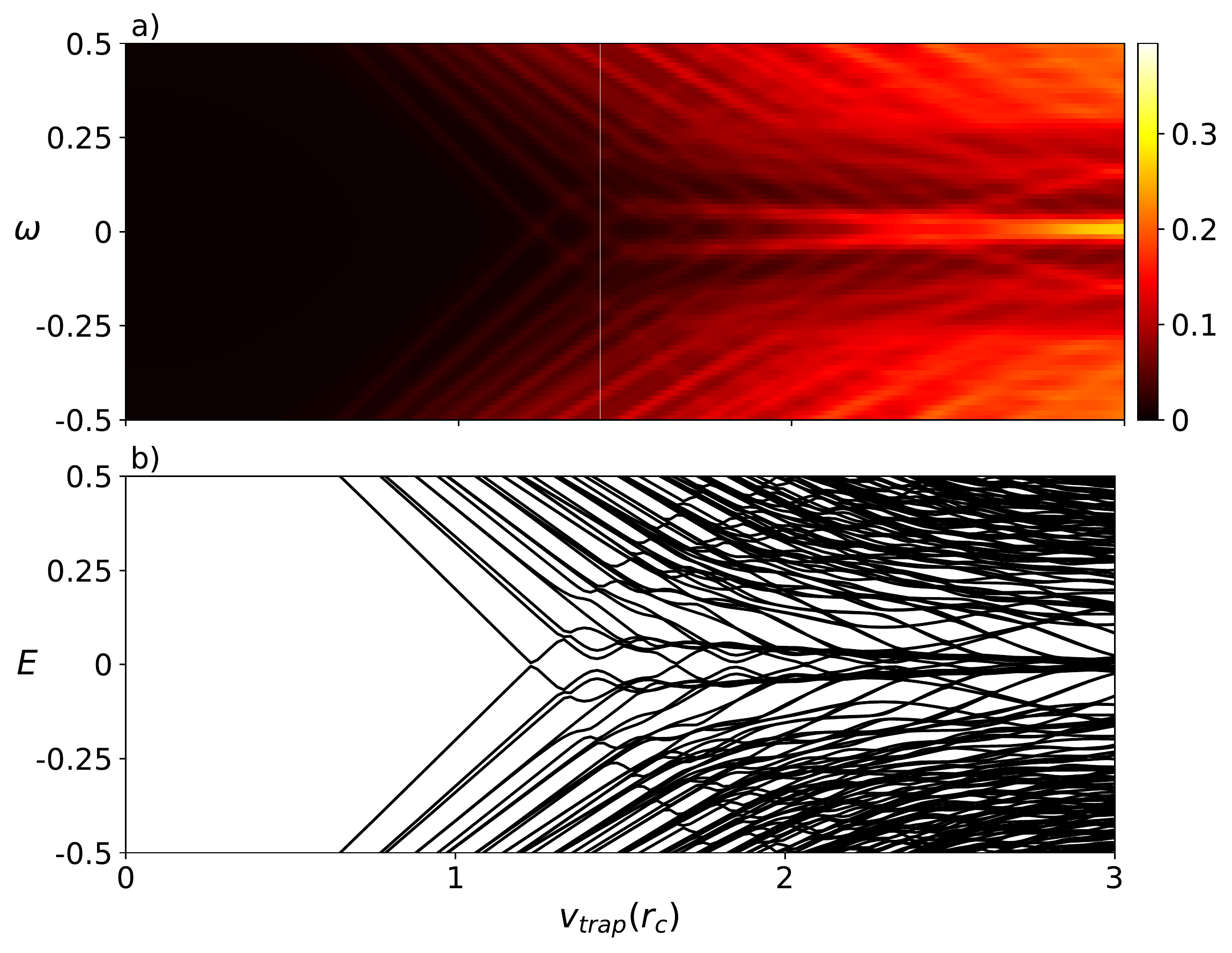}
\caption{(Color online) (a) Density of states, and (b) eigenvalues of the BdG Hamiltonian as a function of $v_{\rm{trap}}(r_c)$ for a $51 \times 51$ square lattice with PBCs. The unit of all values is $\vert m(\bm{k}=0) \vert$.}
\label{fig:pbc}
\end{figure}

We argue that in a second-order TSC, the MCSs-MCFBs transition is actually a crossover in the OBCs system with increasing the HP magnitude: the number of MZMs gradually increases while there is no bulk gap closing as is illustrated in Fig.~\ref{fig:obc}. It can be seen that there are branches of eigenvalues whose values decrease from 0.5 with increasing the HP magnitude although the bulk gap remains open. We find that the proliferation of the MZMs originate from the fact that only 1D gapped boundary modes becomes gapless without the 2D bulk gap closing.

In contrast, in the PBCs system with an increasing HP magnitude, the 2D bulk gap is closed if the potential magnitude at the four corners [$v_{\rm{trap}}(r_c)$] of the 2D TI becomes larger than the insulating gap as shown in Fig.~\ref{fig:pbc}. In this case, by increasing the HP magnitude there are no MZMs because there are no 1D gapped boundary modes to become gapless. We have verified that the same behavior persists for larger systems.
\begin{figure}[t]
\centering
\includegraphics[scale =0.4]{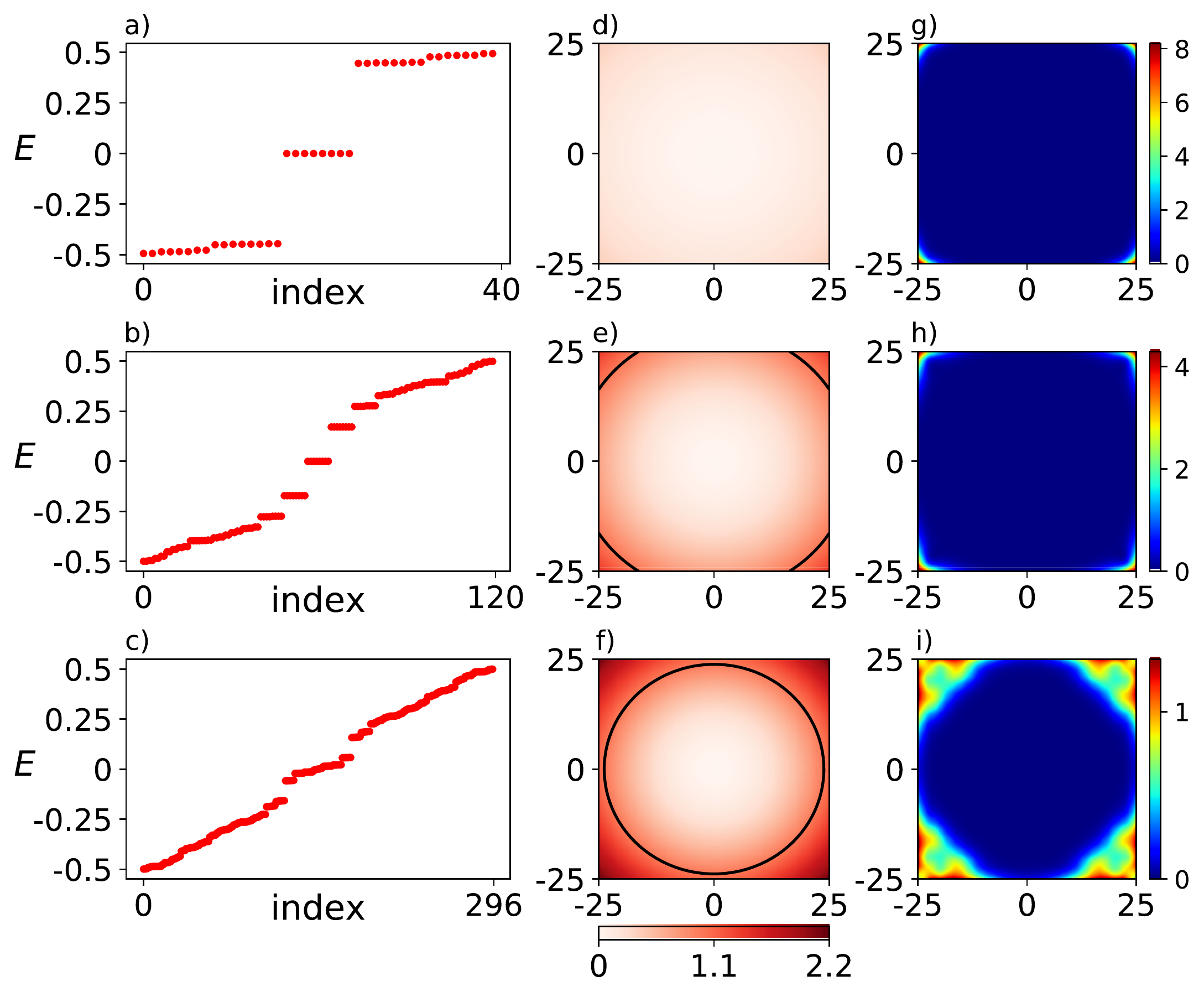}
\caption{(Color online) (a,b,c) Eigenvalues of the BdG Hamiltonian, (d,e,f) two-dimension harmonic potential, and (g,h,i) zero-energy local density of states for a $51 \times 51$ square lattice with OBCs in the presence of three different HP magnitudes $v_{\rm{trap}}(r_c)$ = 0.4 (a,d,g), 1.4 (b,e,h), and 2.2 (c,f,i). Outside the two black circles $v_{\rm{trap}}(r_c) > |m(\bm{k}=0)|$, {\it i.e.}, the HP magnitude is larger than the insulating gap locally. Both the eigenvalues and HP magnitude are given in units of $\vert m(\bm{k}=0) \vert$. The smearing factor in the Lorentzian function for plotting the local density of states is 0.005.}
\label{cross}
\end{figure}

To see the MCSs-MCFBs transition in more detail, we show the real space dependence of the zero-energy local density of states in Fig.~\ref{cross}. In the region outside the black circles of the two lower middle panels of this figure, the HP magnitude is larger than the insulating gap locally, so Figs.~\ref{cross}(b), \ref{cross}(e), and \ref{cross}(h) indicate that MCSs are robust even outside this circle. Moreover, the region outside the circles can host MCFBs, and the additional MZMs originate from only 1D gapped boundary modes that become gapless.
\subsection{Flatness and the origin of MCFBs}
\begin{figure}[!th]
\centering
\includegraphics[scale = 0.48]{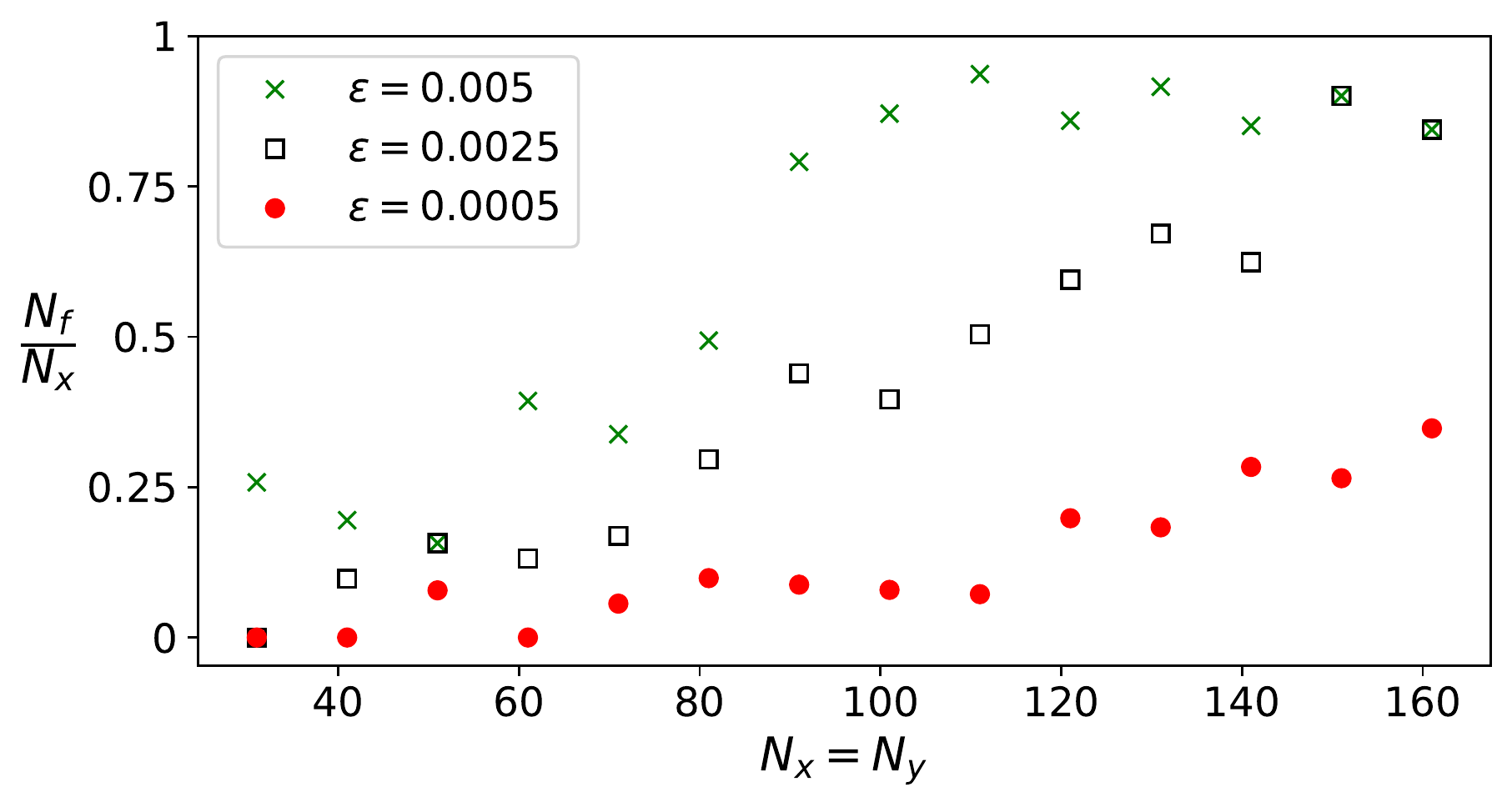}
\caption{(Color online) The ratio between the number of eigenvalues inside the $|E_i| \leq \varepsilon$ interval and the $N_x$ as a function of the size of the boundaries of the lattice for the OBCs at three different values of $\varepsilon$ with $v_{\rm{trap}}(r_c) = 1.1 \vert m(\bm{k}=0) \vert $.}
\label{epsi}
\end{figure}

It should be noted that an exact zero-energy eigenvalue cannot be obtained in the theoretical calculations for a finite system because of finite size effects. Thus, zero energy have to be defined. An eigenvalue can be considered as a zero-energy mode in such a finite system if its magnitude is very small. Therefore, the eigenvalue $E_i$ can be considered as a zero-energy eigenvalue if it satisfies $|E_i| \leq \varepsilon$ condition for a very small $\varepsilon$. By using of this definition for zero-energy modes, we show that as the harmonic trap magnitude is increased, the number of zero-energy eigenvalue increases {\it independent} of the value of $\varepsilon$ for large enough lattice. In Fig.~\ref{epsi}, the ratio between the number of eigenvalues inside the $|E_i| \leq \varepsilon$ interval, say $N_f$, and the number of boundary states, which is proportional to the size of the lattice along $x$ or $y$-direction, is plotted. This figure clearly illustrates that this ratio is increasing in the thermodynamic limit, and a truly flat band appears by increasing harmonic trap magnitude if finite-size effects are suppressed.

\begin{figure}[!th]
\centering
\includegraphics[scale = 0.48]{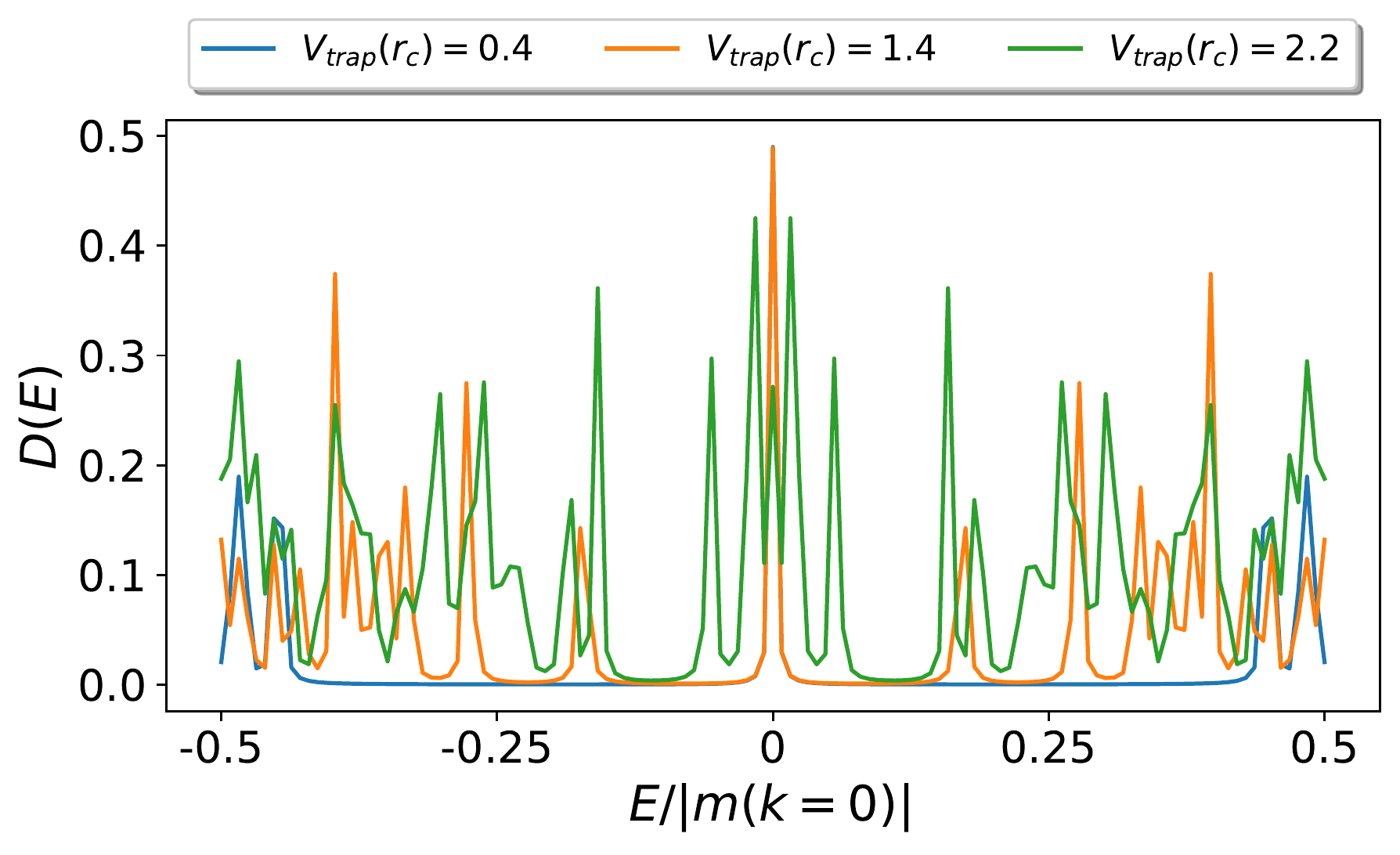}
\caption{(Color online) Density of states as a function of energy in the unit of insulating gap value for the same parameters as the Fig.~\ref{cross}. The bulk gap remains open as harmonic potential magnitude increases in the weak potential limit.}
\label{DOS}
\end{figure}

\begin{figure*}[!th]
\centering
\includegraphics[scale = 0.51]{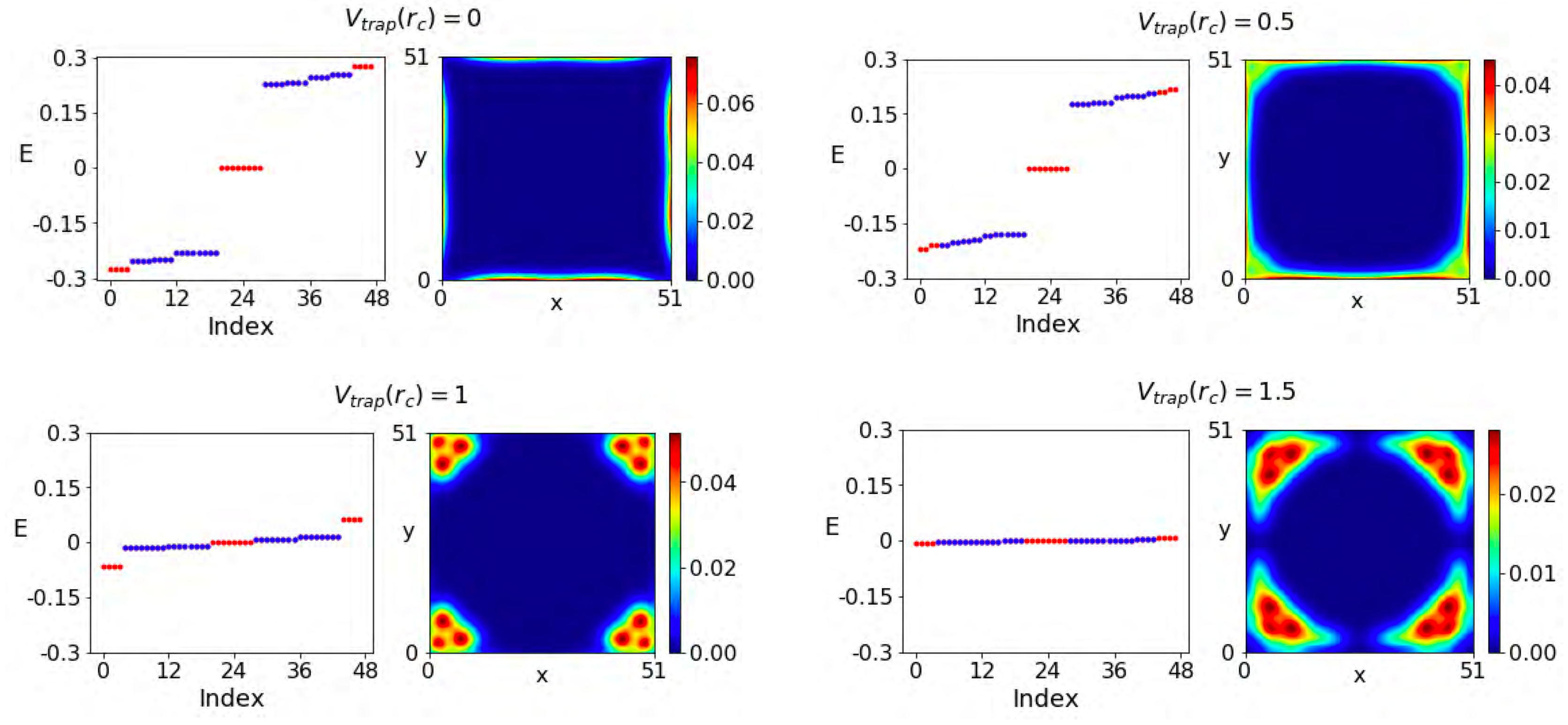}
\caption{(Color online) Eigenvalues of the BdG Hamiltonian and the square of the wave function's amplitude for different values of $v_{\rm{trap}}(r_c)$ and for a $51 \times 51$ square lattice with OBCs.
The square of the wave function's amplitude is plotted only for the eigenvalues that are denoted with blue color. They are gapped boundary states that become gapless by increasing the HP magnitude.}
\label{origin_MFB}
\end{figure*}

In Fig.~\ref{DOS}, we plot the density of states (DOS) for the same parameters as in Fig.~\ref{cross}. It can be clearly seen that the number of Majorana zero modes increases without the bulk gap closing as the harmonic potential magnitude increases in the weak potential limit. The states that appear close to zero energy in this figure are boundary states not bulk states according to Fig.~\ref{origin_MFB}, where the origin of MCFBs is shown. In Fig.~\ref{origin_MFB}, we show that MCFBs originate from only gapped states localized at one-dimensional (1D) boundaries. In this figure we plot the square of the wave function's amplitude associated with only blue points. As the HP magnitude is increased, the energy corresponding to these 1D boundary states become smaller, and they become more localized around the corners. By increasing the HP magnitude, more and more 1D gapped modes become gapless.

\begin{figure}[!b]
\centering
\includegraphics[scale = 0.38]{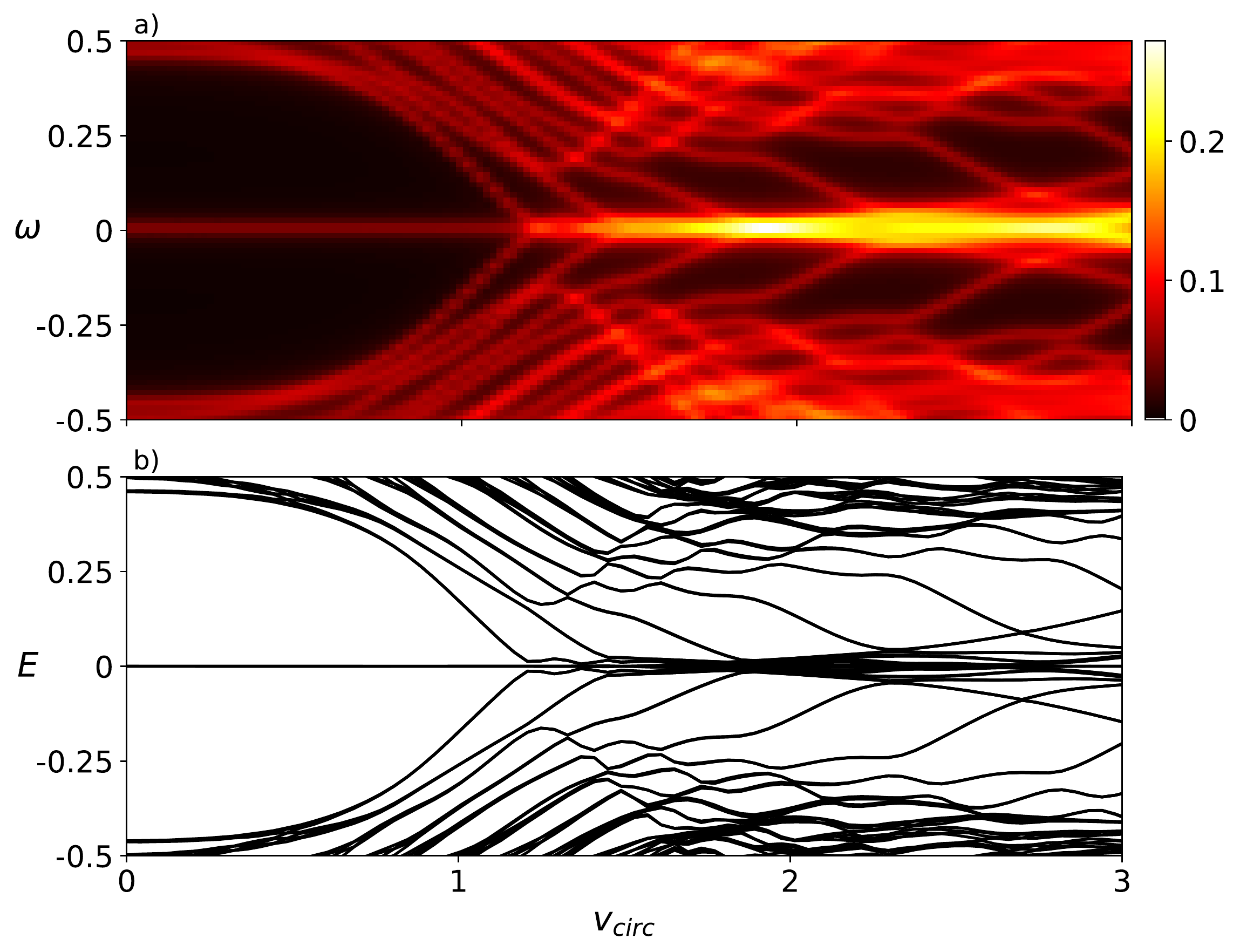}
\caption{(Color online) (a) Density of states, and (b) eigenvalues of the BdG Hamiltonian as a function of $v_{\rm circ}$ for a $51 \times 51$ square lattice with OBCs. The unit of all values is $\vert m(\bm{k}=0) \vert$.}
\label{circ_open}
\end{figure}

\begin{figure}[!b]
\centering
\includegraphics[scale = 0.38]{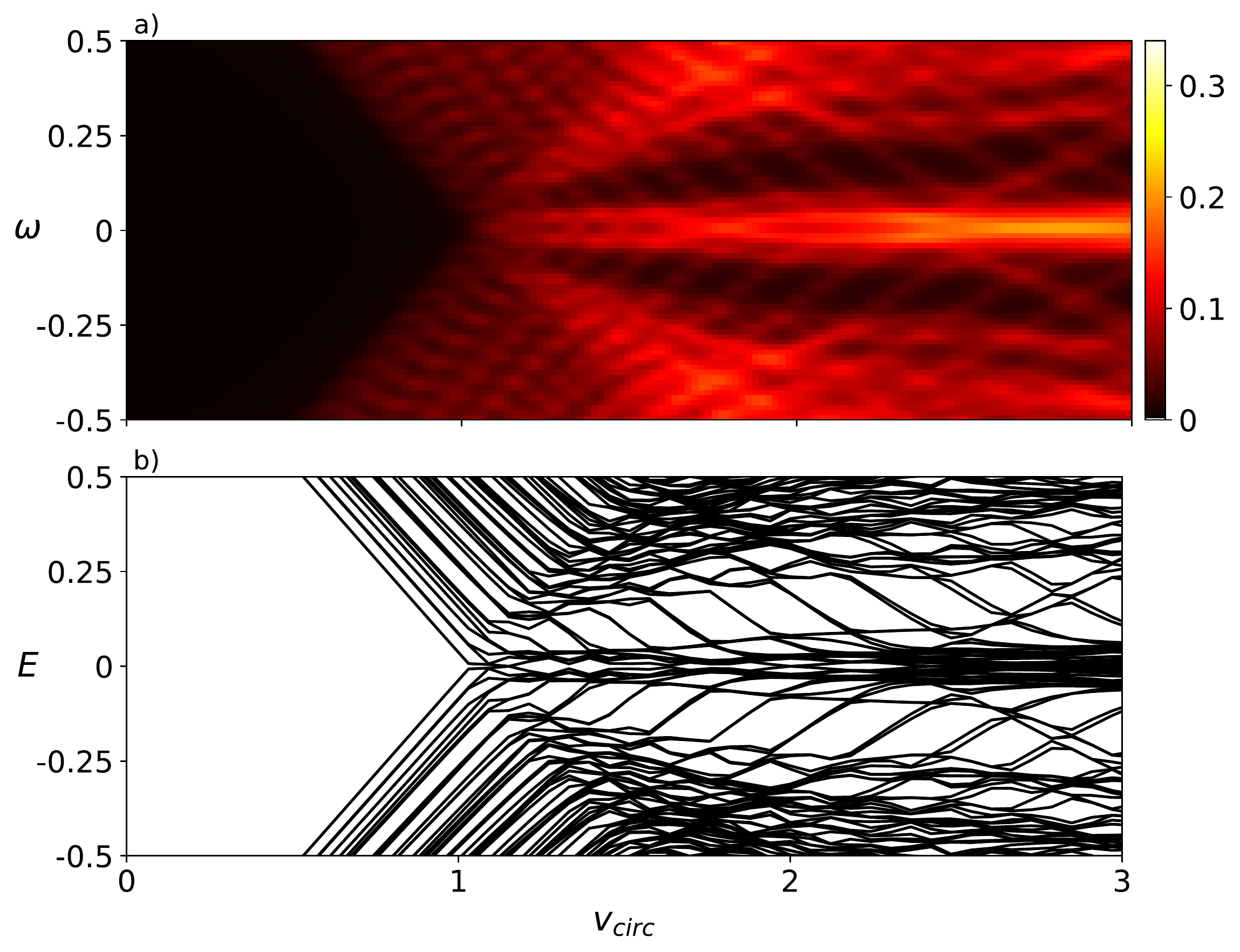}
 \caption{(Color online) (a) Density of states, and (b) eigenvalues of the BdG Hamiltonian as a function of $v_{\rm circ}$ for a $51 \times 51$ square lattice with PBCs. The unit of all values is $\vert m(\bm{k}=0) \vert$.}
  \label{circ_per}
\end{figure}
\subsection{Circular potential}
Let us define the circular potential as $V_i = v_{\rm circ}$ for $(i_x-i^c_x)^2 + (i_y-i^c_y)^2 > R^2$ where $v_{\rm circ}$ is the strength of the circular potential, and $i^c =(i^c_x,i^c_y)$ is the coordinate of the central site of a 2D sample. Here, we take $R = N_x/2$ where $N_x$ is the number of sites along $x$-axis direction. As shown in Figs.~\ref{circ_open} and \ref{circ_per}, the bulk gap closes in both OBCs and PBCs for a circular potential, so there is no crossover and no MCFBs in the presence of a circular potential which is not similar to the case of the harmonic potential.
\section{Discussion}
Let us discuss the difference between conventional MFBs in a $d$-wave SC and MCFBs in our system. In a SC with order parameter symmetry $d_{x^2-y^2}$, it is well known that there are Andreev flat bound states, which appear at the [110] surface. On the other hand, there is no zero-energy bound state at the [100] and [010] surfaces \cite{daido2017majorana, kashiwaya1995origin}. 
For this nodal SC, it is possible to define a 1D partial Brillouin zone by fixing the $(d-1)$-dimensional momentum $\bm{k}_{||}$ at a certain point \cite{sato2011topology, daido2017majorana, yada2011surface, schnyder2011topological}. As long as the pair potential in such a partial Brillouin zone (BZ) is fully gapped, we can define the one-dimensional winding number as the topological invariant of a nodal TSC:
\begin{equation}
\mathcal{W}(\bm{k_{\vert \vert}}) = \frac{1}{2\pi}~ \text{Im} \int dk_{\perp} \partial_{k_{\perp}} \ln \det(q),
\label{wind_num}
\end{equation}
where $k_{\perp}$ is the momentum in a one-dimensional BZ, and $\bm{k}_{||}$ is the momentum parallel to the surface that we consider. Moreover, in Eq.~(\ref{wind_num}), $q(\bm{k})$ is the off-diagonal block of the BdG Hamiltonian in the basis that diagonalizes the chiral symmetry operator such that $\det q(\bm{k})= \big\{ m^2(\bm{k}) + (\Delta(\bm{k}) + i\mu)^2  + \lambda^2 \sin^2 k_x + \lambda^2 \sin^2 k_y \big\} ^2$. 
For a finite winding number, there are $|\mathcal{W}(\bm{k_{\vert \vert}})|$-fold degenerate zero-energy modes at a surface parallel to $\bm{k_{\vert \vert}}$. 
Such highly degenerate surface bound states are known as Majorana flat bands (MFBs) because the energy dispersion is independent of $\bm{k_{\vert \vert}}$ \cite{ikegaya2018symmetry}.

A 2D homogeneous TI in proximity with a $d$-wave SC becomes SC with increasing chemical potential. This transition is a topological phase transition with bulk gap closing since a $d$-wave SC order parameter has line nodes. In this homogeneous system with large enough chemical potential, there are MFBs at the [110] surface because of nontrivial value of winding number. The origin of these MFBs are Andreev bound states \cite{sato2011topology}. However, here we find that in the case of MCFBs not only is there no 2D bulk gap closing in the OBCs system with increasing harmonic potential magnitude, but also the number of MZMs increases only because of the reduction of the energy of the 1D boundary gapped modes. 
Also note that although MCSs are exponentially localized at the corners of the system, MCFBs are localized very close to the four corners in the weak potential limit.

Although MCFBs consist of many very spatially close MZMs, they cannot hybridize and become gapped out. This is in contrast to the fact that MCSs hybridize and annihilate if they are brought close together. 
The reasons for the weak hybridization can naturally be explained by the use of the 1D boundary modes, time-reversal symmetry, and the crossover behavior as follows. Without a confining potential, the wave function of MCSs and 1D gapped boundary modes are orthogonal because they are eigenvectors of the BdG Hamiltonian with different energies.
 The wave functions gradually change with changing the HP magnitude and remain orthogonal since there is no bulk gap closing. Therefore, the crossover makes the MCFBs stable by preserving the orthogonality of these wave functions. In addition, it is known that both Majorana flat bands in a $d$-wave SC, and Majorana corner states in a second-order TSC are protected by time-reversal symmetry \cite{sato2011topology, yan2018majorana}. In our system, in zero and strong potential limits, the system is second-order TSC and $d$-wave SC, respectively. Therefore, the time-reversal symmetry protects the Majorana corner flat bands in the weak potential limit since this phenomenon is a crossover.
\section{Conclusions}
It has been recently proposed that a second-order TSC hosts Majorana corner states (MCSs) localized at the four corners of a 2D TI in proximity with some unconventional high-temperature SCs. This work provides a theoretical framework to make only 1D gapped boundary modes gapless without closing the bulk gap in 2D second-order TSCs. Second-order TSCs always have $(d-1)$-dimensional {\it gapped} boundary modes and $(d-2)$-dimensional gapless modes in $d$-spatial dimensions. Our results suggest that increasing the gradual potential magnitude proliferates the number of MCSs by making {\it only} 1D gapped boundary modes gapless {\it without} closing the bulk gap. This finally leads to a new kind of MFB, namely Majorana corner flat bands (MCFBs). In contrast to the conventional mechanism, MCFBs cannot be characterized by a 1D winding number and only originate from 1D gapped boundary modes that become gapless without closing the bulk gap.

It also has been shown that the presence of time-reversal symmetry during the MCSs-MCFBs crossover prevent MCSs from hybridizing and protect MCFBs. There might be a topological invariant for this particular inhomogeneous system to characterize the crossover, but finding that in real-space is a mathematically challenging task and beyond the scope of our present paper.

In experiment, if one can observe the MCSs in second-order TSCs, such as a hybrid structure of TI and an unconventional SC or even cold atom systems, one can also observe the MCSs-MCFBs crossover and MCFBs with adding a gradual potential such as HP. More precisely speaking, for having a hybrid structure of TI and an unconventional SC, a monolayer of WTe$_2$ as a high-temperature topological insulator \cite{qian2014quantum, wu2018observation} can be exploited in proximity to a $d$-wave high-temperature cuprate superconductor. In this hybrid solid-state systems, some potentials can be applied by using of electron double-layer transistors (EDLTs) \cite{li2016controlling, ahn2006electrostatic, weisheit2007electric}, which can be used to apply an electric field to the sample. 
Normally the potentials generated by the EDLT techniques have sharp edges. Nevertheless, there might be some possibilities in the future or other advanced techniques to control the sharpness of this trapping potential in order to approach a harmonic potential.
This work might open up new prospects for realizing MZMs and its applications in quantum computation and information, which will be left for future investigations.
\vskip0.2in
\begin{acknowledgments}
We would like to acknowledge R.~Boyack for helpful comments. We also acknowledge enlightening discussions with T.~Nojima concerning experimental realizations of the
EDLT techniques and with L.~LeBlanc concerning experimental realizations in cold atom systems. This work was supported in part by the Natural Sciences and Engineering Research Council of Canada (NSERC). The calculations were partially performed by the supercomputing systems SGI ICE X at the Japan Atomic Energy Agency. Y.~N. was partially supported by JSPS KAKENHI Grant Number 18K11345 and 18K03552, the ``Topological Materials Science'' (No. JP18H04228) KAKENHI on Innovative Areas from JSPS of Japan. 
\end{acknowledgments}
\appendix*
\section{Majorana Condition}
Although the BdG Hamiltonian in Eq.~(\ref{hami_1}) of the main paper is a $8N \times 8N$ matrix, by changing the basis it can be written as a block diagonal matrix where each block is a $4N \times 4N$ matrix and $N$ is the number of lattice sites. In this section, it is shown that the zero modes of each of these $4N \times 4N$ blocks meets the Majorana conditions. The Hamiltonian can be rewritten as, 
\begin{align}
{\cal H} = \frac{1}{2} \Psi^{\dagger} \hat{H}  \Psi,
\end{align}
here
$\bm{\psi}^{\dagger}_i=(
c^{\dagger}_{i,a,\uparrow},
c^{\dagger}_{i,b,\uparrow},
c^{\dagger}_{i,a,\downarrow},
c^{\dagger}_{i,b,\downarrow},
c_{i,a,\uparrow},
c_{i,b,\uparrow},
c_{i,a,\downarrow},
c_{i,b,\downarrow})$
and,
\begin{align}
\hat{H} &= \left(\begin{array}{cc}
H_n & \Delta \\
\Delta^{\dagger} & -H_n^{\ast}
\end{array}\right),
\label{hamii}
\end{align}
where
\begin{align}
H_n = \left(\begin{array}{cc}
H_1 & 0 \\
0 & H_2
\end{array}\right),
\hspace{0.5cm}
\Delta = 
\left(\begin{array}{cc}
0 & -\Delta \\
\Delta & 0
\end{array}\right).
\end{align}
The BdG equations are 
\begin{align}
\hat{H} \left(\begin{array}{c}
\vec{u}_1 \\
\vec{u}_2 \\
\vec{v}_1 \\
\vec{v}_2
\end{array}\right)  &= E 
\left(\begin{array}{c}
\vec{u}_1 \\
\vec{u}_2 \\
\vec{v}_1 \\
\vec{v}_2
\end{array}\right) ,
\end{align}
or
\begin{align}
H_1 \vec{u}_1 - \Delta \vec{v}_2 &= E \vec{u}_1 \\
H_2 \vec{u}_2 + \Delta \vec{v}_1 &= E \vec{u}_2 \\
\Delta^{\ast} \vec{u}_2 - H_1^{\ast} \vec{v}_1 &= E \vec{v}_1 \\
-\Delta^{\ast} \vec{u}_1 -H_2^{\ast} \vec{v}_2 &= E \vec{v}_2.
\end{align}
So, we have a $4N \times 4N$ matrix:
\begin{align}
\left(\begin{array}{cc}
H_1 & - \Delta \\
-\Delta^{\ast}  & -H_2^{\ast}
\end{array}\right)
\left(\begin{array}{c}
\vec{u}_1 \\
\vec{v}_2
\end{array}\right) &= E
\left(\begin{array}{c}
\vec{u}_1 \\
\vec{v}_2
\end{array}\right).\label{eq:4n}
\end{align}
Then, it can be shown that,
\begin{align}
-H_1^{\ast} \vec{u}_1^{\ast} + \Delta^{\ast} \vec{v}_2^{\ast} &= -E \vec{u}_1^{\ast} \\
-H_2^{\ast} \vec{u}_2^{\ast} - \Delta^{\ast} \vec{v}_1^{\ast} &= -E \vec{u}_2^{\ast} \\
-\Delta \vec{u}_2^{\ast} + H_1 \vec{v}_1^{\ast} &= -E \vec{v}_1^{\ast} \\
\Delta \vec{u}_1^{\ast} + H_2\vec{v}_2^{\ast} &= -E \vec{v}_2^{\ast}
\end{align}
or
\begin{align}
\hat{H}
\left(\begin{array}{c}
\vec{v}_1^{\ast} \\
\vec{v}_2^{\ast} \\
\vec{u}_1^{\ast} \\
\vec{u}_2^{\ast}
\end{array}\right) &= -E 
\left(\begin{array}{c}
\vec{v}_1^{\ast} \\
\vec{v}_2^{\ast} \\
\vec{u}_1^{\ast} \\
\vec{u}_2^{\ast}
\end{array}\right).
\end{align}
Therefore, if $(\vec{u}_1^i,\vec{u}_2^i,\vec{v}_1^i,\vec{v}_2^i)$ is an eigenstate with energy $E_i$, 
then $(\vec{v}_1^{i \ast},\vec{v}_2^{i \ast},\vec{u}_1^{i \ast},\vec{u}_2^{i \ast})$ is an eigenstate with energy $-E_i$ which should be expected because of particle-hole symmetry.
This means that if $E = 0$, we have the following relations,
\begin{align}
\vec{u}_1^i = \vec{v}_1^{i \ast} \label{eq:relation1}
\hspace{1.5cm}
\vec{u}_2^i &= \vec{v}_2^{i \ast}.
\end{align}
In other words, if we have $(\vec{u}_1^i,\vec{v}_2^i)$ with the energy $E_i$, we can obtain eigenstates $(\vec{u}_2^i,\vec{v}_1^i) = (\vec{v}_2^{i \ast},\vec{u}_1^{i \ast})$ with the energy $-E_i$.
It means by finding only the eigenvalues and eigenstates of the $4N \times 4N$ Hamiltonian in Eq.~(\ref{eq:4n}), one can find all eigenvalues and eigenstates of the $8N \times 8N$ Hamiltonian in Eq.~(\ref{hamii}) much more easily due to the particle-hole symmetry. In the rest of this section, we will show that all the zero modes of the $4N \times 4N$ Hamiltonian satisfy Majorana conditions.

The unitary matrix that diagonalizes the matrix $\hat{H}$ is expressed as 
\begin{align}
\hat{U} = \left(\begin{array}{cc}
\hat{u} & \hat{v}^{\ast} \\
\hat{v} & \hat{u}^{\ast} 
\end{array}\right),
\hspace{0.4cm}
\hat{H} = \hat{U} \left(\begin{array}{cc}\hat{E} & 0 \\
0 & -\hat{E}
\end{array}\right)\hat{U}^{\dagger}
\end{align}
where
\begin{align}
\hat{u} \equiv \left(\begin{array}{ccc}
\vec{u}_1^1 & \cdots  & \vec{u}_1^N \\
\vec{u}_2^2 & \cdots & \vec{u}_2^N
\end{array}\right),
\hspace{0.4cm}
\hat{v} \equiv \left(\begin{array}{ccc}
\vec{v}_1^1 & \cdots  & \vec{v}_1^N \\
\vec{v}_2^2 & \cdots & \vec{v}_2^N
\end{array}\right).
\end{align}
So, 
\begin{align}
{\cal H} &= \frac{1}{2} \Psi^{\dagger} \hat{U} \left(\begin{array}{cc}\hat{E} & 0 \\
0 & -\hat{E}
\end{array}\right)\hat{U}^{\dagger}
  \Psi
  \\&
  =\frac{1}{2} \sum_i (E_i \gamma_i^{\dagger} \gamma_i  -E_i \gamma_i \gamma_i^{\dagger}),
\end{align}
where
\begin{align}
\left(\begin{array}{c}\vec{\gamma} \\
\vec{\gamma^{\dagger}} 
\end{array}\right) &= 
\hat{U}^{\dagger} \left(\begin{array}{c}\vec{c} \\
\vec{c^{\dagger}} 
\end{array}\right)
\end{align}
or
\begin{align}
\gamma_i &= \sum_l \left[ [\hat{u}]_{li}^{\ast} c_l + [\hat{v}]_{li}^{\ast}  c_l^{\dagger} \right]
\\
\gamma_i^{\dagger} &= \sum_l \left[ [\hat{u}]_{li}c_l^{\dagger}  + [\hat{v}]_{li}  c_l\right].
\end{align} 
It is obvious that $\gamma_i^{\dagger} \neq \gamma_i$. However, we have two special conditions for only zero-eigenvalues which is Eq.~(\ref{eq:relation1}). Using them we will have,
\begin{align}
\gamma_i^{\dagger} &= \sum_l \left[ [\hat{v}]_{li}^{\ast} c_l^{\dagger}  + [\hat{u}]_{li}^{\ast}   c_l \right] = \gamma_i.
\end{align}
Therefore, the zero-energy eigenstates are all Majorana.

\bibliography{my_bib}

\begin{thebibliography}{61}
\bibitem [{rem()}]{remark1}%
  \BibitemOpen
  \href@noop {} {\ }\bibinfo {note} {This assumes translational invariance,
  which precludes the realistic possibility that the order parameter changes
  near the surface of a material compared to the bulk, for
  example.}\BibitemShut {Stop}%
\end{thebibliography}
\end{document}